\documentclass[11pt,a4paper]{article} % 10pt is ignored!
\pdfoutput=1
\usepackage{jheppub}
\usepackage{graphicx}
\usepackage{amsmath,amssymb}
\usepackage{slashed}

\newcommand{\ct}{{\cos\theta}}
\newcommand{\ctv}{{\cos\theta^V}}
\newcommand{\Av}{{\mathcal{A}^V_{ll}}}
\newcommand{\ctll}{{\cos\theta_{ll}}}

\newcommand{\nn}{\nonumber}
\newcommand{\ba}{\begin{equation}}
\newcommand{\ea}{\end{equation}}
\newcommand{\be}{\begin{eqnarray}}
\newcommand{\ee}{\end{eqnarray}}

\title{Spin before mass at the LHC}
\author{Tom Melia} 

\affiliation{Rudolf Peierls Centre for Theoretical Physics, 1 Keble Road, University of
   Oxford, UK}
\emailAdd{ t.melia1@physics.ox.ac.uk}

\abstract{
What can be said about the spin of new particles without knowing their mass during an initial discovery phase at CERN?
We consider this question in a topology where mass measurement is particularly difficult, $pp\to Y\overline{Y}\to lX\overline{l}\,\overline{X}$,
and introduce two new variables $\ctv$ and $\Av$ which we prove are both independent of the mass of $X$. The variable $\ctv$ approximates
the polar production angle of $Y$, and we find that it possesses greater statistical power in determining the spin of this particle 
than a previous, related variable, $\ctll$ \cite{Barr:2005dz}. $\Av$ is an asymmetry which can provide information about the couplings of
a spin half $Y$. Because these variables can be used from the outset, without any knowledge of the masses of the new particles,
we find here that it is possible to reverse the usual `mass before spin' determination timeline.
}
\begin{document} 

\maketitle

\section{Introduction}

To discover physics beyond the standard model at the LHC would be a great success. If new particles are
created at CERN, their masses, spins and other quantum numbers will illuminate the
details of the theory describing nature at the TeV scale.

\begin{figure}[t]
\begin{center}
\includegraphics[width=8cm]{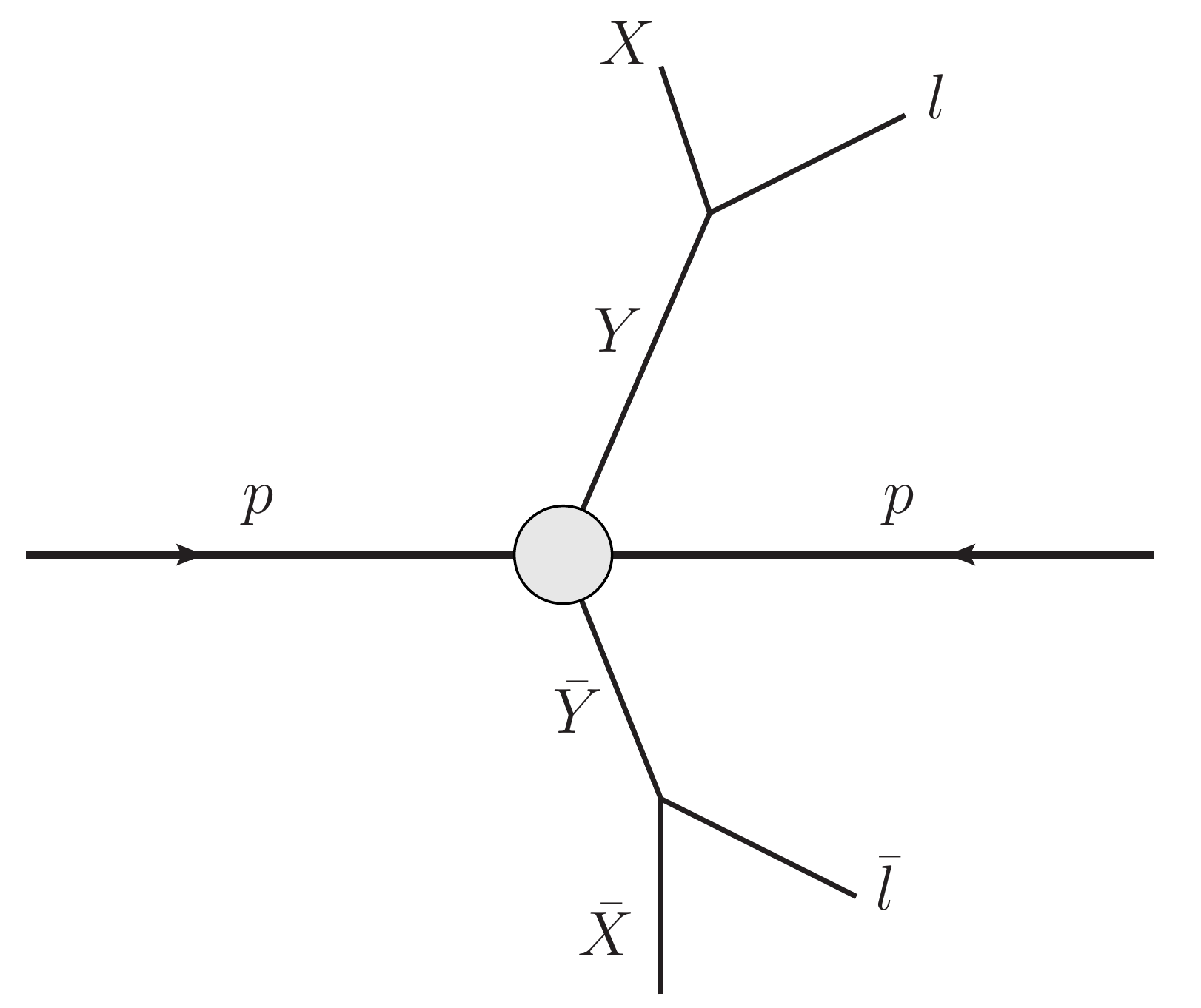}
\caption{The event topology $pp\to Y\overline{Y}\to l X \overline{l}\, \overline{X}$.}
\label{topo}
\end{center}
\end{figure}

In this paper, we study a method of spin determination in the topology shown in figure~\ref{topo}: a new particle-antiparticle
pair $Y\overline{Y}$ is produced, followed by the decays $Y\to lX$ and $\overline{Y}\to\overline{l}\,\overline{X}$ where $l$ is either an 
electron or a muon, and $X$ and $\overline{X}$ are the lightest of the new particle states, which we assume are stable and 
invisible to the detector. We investigate the use of observables which do not take as input the mass spectrum of $Y$ and $X$
and show that the two introduced here are formally independent of the mass of $X$. The first is related to Barr's $\ctll$ \cite{Barr:2005dz},
 which approximates the polar production angle, $\ct$, of $Y$ with respect to the beam axis in the $Y\overline{Y}$ centre of mass frame.
  We find an improvement on $\ctll$'s statistical power in determining the spin of $Y$. The second is an asymmetry which picks
 up on a term which can be present in the production matrix element squared for a spin half $Y$, $|\mathcal{M}|^2\propto \pm \ct$.
 We will discuss the possibility of using this asymmetry to provide information about the couplings of $Y$, something which is difficult
 to do using polar angle distributions alone.

The problems that come with having two invisible particles in the final state of an event are well documented. Because mass measurement is 
usually `easier' than spin measurement, in the sense that fewer events are required to obtain a reasonably accurate result, methods of spin
determination are often happy to make the assumption that the mass spectrum of the theory is known. 
We do not go into detail here as to the methods which can be employed at hadron colliders to measure mass, and instead refer the reader to a recent review 
\cite{Barr:2010zj}.
It is true, however, that the event
topology we are considering is an example in which mass determination is particularly difficult \cite{Konar:2009wn}. 
 This is down to not having two or
more visible particles in the decay chain of $Y$ -- there are no edges or endpoints which could come from the invariant mass distributions
of such particles to provide relations between the new state masses and help to pin them down \cite{Barr:2010zj}. Nor is there an experimentally stark
$M_{T2}$ `kink'  \cite{Lester:1999tx, Barr:2003rg, Cho:2007qv} which would point directly to the mass of both particles. It would require many events 
to determine
even the mass scale of the new physics, or else rely on mass measurements from elsewhere. At any rate, it is interesting to ask the question: what can we say about spin without any 
knowledge of the particle masses? 

Likelihood analysis techniques for spin determination use the complete event information and so can make statistically optimal statements, but their
implementation becomes extremely complicated if the masses are unknown and need to be scanned over. Methods based
on the total cross section \cite{Datta:2005vx} can also be very powerful, as can those which use kinematic reconstruction \cite{Cho:2008tj,Cheng:2010yy,Horton:2010bg}, 
but they too require at least the masses to be known before they become useful. 
 We shall
investigate spin determination during an initial discovery phase, before these two techniques become viable, simply by studying the angular distributions
of visible particles in this topology -- these are affected by the particular spin configuration at hand \cite{Barr:2005dz, Chen:2010ek, MoortgatPick:2011ix, Buckley:2008eb}. 
We shall find that the properties of the observables
introduced here make them ideal for use in a `no mass input' approach and this leaves us in a position where it actually becomes
possible to determine the spin of a new particle before we know its mass.

In section~\ref{sec:pair} we discuss how the spin of $Y$ can
affect its production, identifying features which we can try to pick up on through the lepton momenta -- this will also serve to introduce the
set-up used in this paper.
In section~\ref{sec:decay} we define the variable $\ctv$, which is similar to $\ctll$, and the asymmetry $\Av$, which can be present
if $Y$ is spin half, and which gives insight into the vector-axial nature of its coupling. The interesting mass independence properties
of both variables are studied here too. We demonstrate their use in determining the spin
of $Y$ in section~\ref{sec:coll} and present conclusions in section~\ref{sec:conc}.

\section{Pair-production of $Y\overline{Y}$}
\label{sec:pair}

For the present, we shall pretend that $Y$ and $\overline{Y}$ are produced, do not decay and 
that we can measure their four-momentum perfectly. The following section will deal with the
real-world problem of not being able to make this measurement when the final state cannot be reconstructed.

\begin{figure}[t]
\begin{center}
\includegraphics[width=8cm]{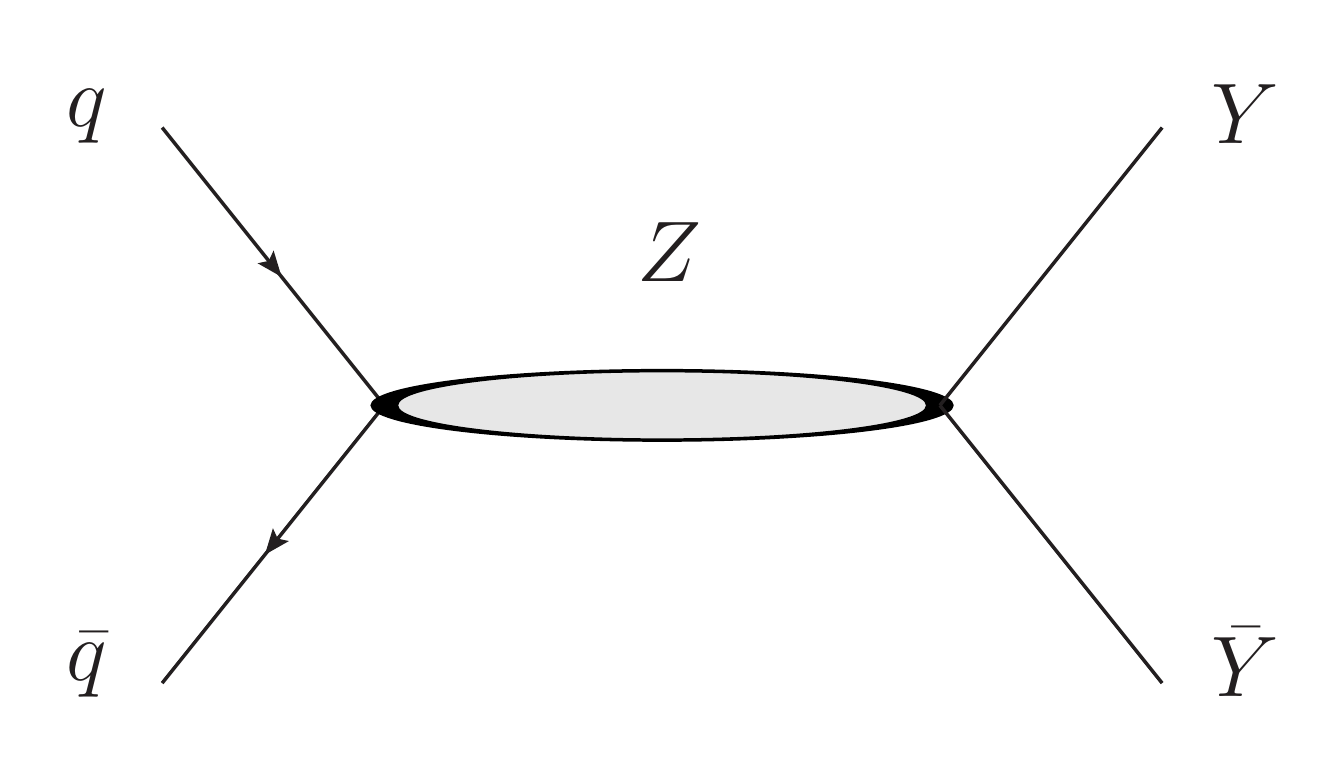}
\caption{Production of $Y\overline{Y}$ by a spin one particle, $Z$, in the $s$-channel.}
\label{prod}
\end{center}
\end{figure}

We have chosen in this paper to illustrate the use of our variables by considering the $s$-channel production of $Y$ and $\overline{Y}$
by a spin one particle $Z$, as shown in figure~\ref{prod}, where $Y$ is either a spin zero or a spin half particle. 
The particle $Z$ could be either the standard model $Z^0$ weak boson or it could be 
some new spin one state.
We aim to discriminate between these two spin assignments where the $Y$s are produced according to distributions
which go something like  $\sim\sin^2\theta$ and 
$\sim (1+\cos^2\theta)$ respectively.  We make the assumption that one can neglect all production
diagrams other than the one in figure~\ref{prod} -- we shall discuss how this situation can arise shortly -- and continue in a
model independent fashion. 
%A study of spin one $Y$ is inconsistent within this framework, as $t$-channel diagrams are necessary in order
%to stop the cross section growing without bound, and we will not consider this scenario\footnote{The variables introduced could just as 
%easily identify a spin one $Y$, so long as the distribution of its polar production angle is distinct from the other spin configurations. This may
%not be the case, however.}. 
We allow for the couplings between $Z$ and $Y$ to be those of the most general, renormalisable interaction:
$CZ^\mu(Y^*(\partial_\mu Y) - (\partial_\mu Y^*) Y )$ for a spin zero $Y$ and $\overline{Y}\slashed{Z}(V+A\gamma_5)Y$ for a spin half $Y$,
where $C$, $A$ and $V$ are the coupling constants. Because we are considering an early discovery phase at the LHC when the cross
section for such a process won't be accurately known, we shall be comparing normalised distributions of variables. Given this, in the spin zero case there is no
room for manoeuvre -- the spin contribution is fixed as there is just an overall factor of the coupling constant, $C$. On the other hand, for the
spin half case, there is the relative size of $V$ and $A$, the vector-like and axial-like couplings of $Y$ to $Z$, which is a free parameter in this set-up.

Under what conditions does the diagram of figure~\ref{prod} dominate the production of $Y\overline{Y}$? The importance of other diagrams, especially
$t$-channel ones, must be minimal, since these can serve to make equal the angular distributions for spin zero and spin half $Y$ \cite{Battaglia:2005zf}. 
Figure~\ref{prod} automatically depicts the dominant production mechanism in the case
 where $M_Z>2M_Y$ so that the particle $Z$, some new state, is resonantly produced.
Alternatively, one could simply imagine that some symmetry of the theory describing these particles forbids any $t$-channel 
diagram\footnote{For example in \cite{Barr:2005dz} $Z=Z^0$ and so $M_Z<2M_Y$ and there is no resonance. Here the $Y$s are sleptons and
conservation of lepton and baryon number forbids any tree level $t$-channel diagrams.}.
Because the former situation is more general, we choose to select
some example masses for $Y$ and $Z$ so that $Z$ can be resonantly produced, and use these values 
to illustrate the various steps along the way. We take $M_Y=800\,$GeV and $M_Z=2.5\,$TeV, and we
shall return in section~\ref{sec:coll} to discuss how different choices for these masses affect our results (and will find that so long as
$Y$ and $\overline{Y}$ are reasonably well boosted in the LAB frame, which can happen for both $M_Z\gtrless2M_Y$, then this is all we need to 
leave our conclusions unchanged). We shall also present results 
for the case when $Z$ is the standard model $Z^0$.

The two different spin assignments provide two different matrix elements to evaluate. We choose to do so in the centre of mass frame of the $Y\overline{Y}$ pair, where we write $Y$'s momentum as
\be
p^{\text{CM}}_{Y} =  \gamma M_Y \left(\begin{array}{c}
1\\
0\\
\beta\,\sin\theta\\
\beta\,\ct
\end{array}\right)\,,
\ee
where $\beta$ is the velocity of $Y$ and $\gamma=1/\sqrt{1-\beta^2}$. In terms of these variables, the matrix element squared, summed over final particle spin and averaged over initial quark spin and colour is
\be
\overline{\left|\mathcal{M}\right|^2} = \mathcal{N} \; F_+ \; P(\gamma) \; \text{SPIN}(\beta,\ct)\,,
\label{mat}
\ee
where
\be
P(\gamma) = \frac{\gamma^4}{(4\,\gamma^2-M_Z^2/M_Y^2)^2+(\Gamma_ZM_Z/M_Y^2)^2}
\ee
shows the effect of the $Z$ propagator with a Breit-Weigner width $\Gamma_Z$, $\mathcal{N}$ is a numerical constant and
$F_+$ is given in terms of the vector and axial couplings
of $Z$ to the quarks
\be
F_+=(f_q^V)^2+(f_q^A)^2\,.
\ee
Now we turn to the important difference. For a spin zero $Y$
\be
\text{SPIN}(\beta,\ct) = C^2\,\beta^2 \left( 1- \cos^2\theta \right)\,,
\label{spinz}
\ee
whereas for a spin half $Y$
\be
\text{SPIN}(\beta,\ct) = V^2\left(2+\beta^2(\cos^2\theta-1)\right) + A^2 \,\beta^2 \,\left(1+\cos^2\theta \right) \pm 8 \frac{f^A_qf^V_q}{F_+} V A \,\beta\,\ct\,.
\label{spinh}
\ee
We see an interesting final term in equation~\eqref{spinh}, asymmetric in $\ct$, which arises from an $\epsilon^{\mu\nu\rho\sigma}$ term coming out of the Dirac matrix algebra. This term is positive if one takes $\theta=0$ in the direction of the quark, and is negative if it is taken in the direction of the antiquark. Non-zero vector and axial couplings are needed at both the $Zq\overline{q}$ and the $ZY\overline{Y}$ vertices for this term
to be present in the first place.
At this stage then we discuss the coupling of $Z$ to the quarks, through $f_q^V$ and $f_q^A$. The overall factor of  $F_+$ in equation~\eqref{mat} will 
not affect the normalised distributions we are interested in and so the only place these couplings matter is in determining the relative size of the asymmetric term.  We investigate by considering the case where $ {f^A_qf^V_q}/{F_+}\sim\mathcal{O}(1)$ and find it useful to just take the
values of these couplings as we would were $Z$ the standard model $Z^0$. This way, when we later set the mass and width of $Z$ to those of $Z^0$, 
this particle couples as the true standard model particle to the quarks.

\begin{figure}[t]
\begin{center}
\includegraphics[width=8.7cm]{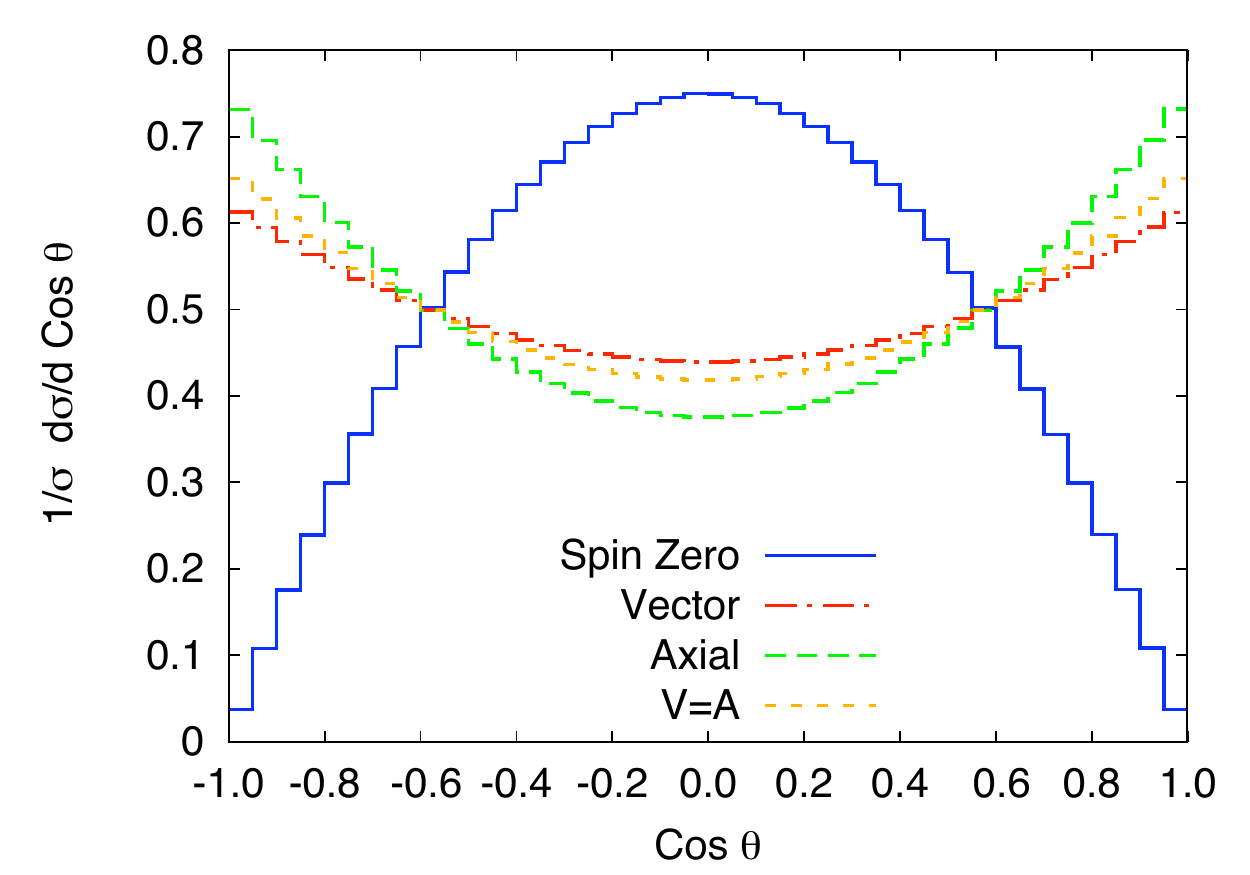}
\includegraphics[width=6.3cm]{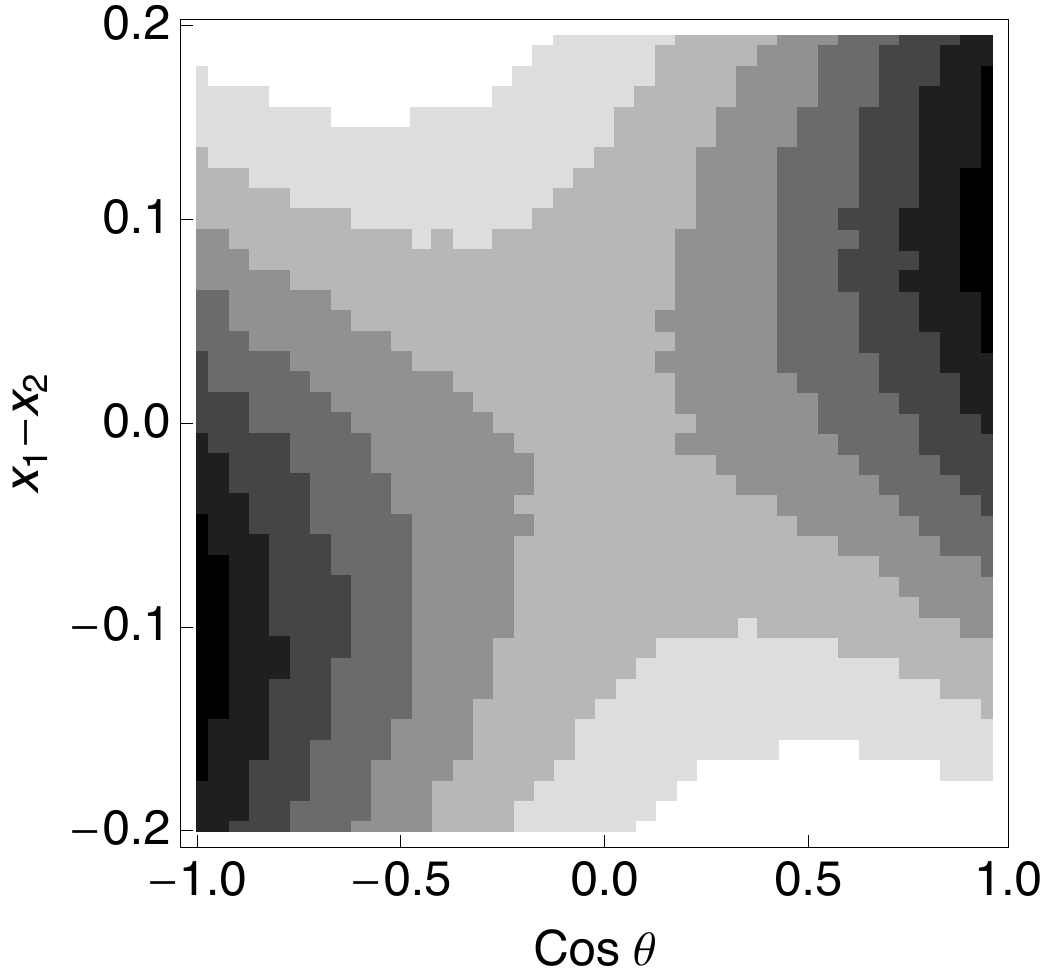}
\caption{Two plots useful for spin determination if we were able to reconstruct the final state and thus know the momentum of $Y$. The left hand diagram shows the 
normalised distribution of $\ct$, for the case when $Y$ is spin zero and the case when $Y$ is spin half with a purely vector coupling ($A=0$), a purely axial coupling ($V=0$), and mixed couplings ($V=A=1$ is chosen). In the right hand diagram, the $V=A$ case is also projected along the $x_1-x_2$ direction, revealing the `hidden' asymmetry. Here, darker areas mean more events and the normalisation is arbitrary. We use the example masses for $Y$ and $Z$ described in the text. }
\label{cos}
\end{center}
\end{figure}

It remains to integrate these expressions over phase space and convolute this parton level process with the parton distribution functions \cite{Pumplin:2002vw}
for the proton. We do this for $pp$ collisions at $14\,$TeV, using the values of our example masses outlined above. 
The resulting normalised differential cross sections are shown in the left hand diagram of figure~\ref{cos} for four different cases: a spin zero $Y$; a spin half $Y$ with 
a purely vector coupling (that is, setting $A=0$ in equation~\eqref{spinh}); a spin half $Y$ with purely axial couplings (setting $V=0$ in equation~\eqref{spinh}); and a spin half $Y$ with mixed couplings, where we choose for illustration $V=A=1$. Note the lack of asymmetry even when 
the $VA\ct$ term is present.

The prospect of distinguishing between a spin zero and a spin half $Y$ looks promising using the variable $\ct$ (remember we are
imagining that we can measure $Y$ and $\overline{Y}$'s momenta and so know how to boost to their centre of mass frame), and we may even be 
hopeful about telling the difference between an axially coupling $Y$ and the other spin half $Y$s. It is, however, disappointing that the asymmetry
has been lost, especially as this could give a handle on differentiating between the purely vector and the 
mixed axial-vector cases -- something which is difficult to do using $\ct$ alone. We should not be surprised that it is lost -- the symmetry of the initial $pp$
state of course prevents such an asymmetry being present in the distribution of $\ct$. To understand this in terms of equation~\eqref{spinh}, if
we choose to measure $\ct$ with respect to the direction of motion of the right-moving proton (which we call proton 1), then we
are just as often defining $\theta=0$ with respect to the direction of motion of a quark as we are an anti-quark.

Although this asymmetry has been `hidden', it is possible to measure it still -- to do so we have to introduce an asymmetry of our
own into the analysis. This can be done by distinguishing between events which are boosted in the positive $z$ direction and those
boosted in the negative $z$ direction. To see how this can reintroduce the asymmetry, consider the case when the $Y\overline{Y}$ rest frame 
is boosted in the positive $z$ direction in the LAB frame. This means that $x_1>x_2$, where $x_1$ and $x_2$ are the momentum fractions
of the partons coming from proton 1 and proton 2 (left-moving) respectively, and are given by 

\be
x_1 &=& \frac{1}{\sqrt{S}} (E_Y+E_{\overline{Y}}+p^z_{Y}+p^z_{\,\overline{Y}}) \nn \\
x_2 &=& \frac{1}{\sqrt{S}} (E_Y+E_{\overline{Y}}-p^z_{Y}-p^z_{\,\overline{Y}})\,,
\ee
where $\sqrt{S}$ is the centre of mass energy of the proton-proton system. For $x_1>x_2$, the parton distribution functions yield a higher
probability for parton 1 to be a quark and parton 2 to be an antiquark, rather than the other way round. Overall there is a
preference for the + sign in equation~\eqref{spinh} and we see this in the right-hand diagram of figure~\ref{cos} which plots the normalised
cross section in the ($\ct,x_1-x_2$) plane. We can now define an asymmetry as
\be
\mathcal{A}_{ll} = \frac{N^{x_1>x_2}_{\ct>0}+N^{x_1<x_2}_{\ct<0}-N^{x_1>x_2}_{\ct<0}-N^{x_1<x_2}_{\ct>0}}{N_{\text{TOT}}}
\ee
where $N^{x_1\gtrless x_2}_{\ct\gtrless 0}$  means the number of events with $x_1\gtrless x_2$ and $\ct\gtrless 0$ and $N_{\text{TOT}}$ is the total number
of events.

Both $\ct$ and $\mathcal{A}_{ll}$ can tell us something about the spin of $Y$ and in the latter case, something about $Y$s coupling too.
We shall now see whether we can construct similar variables using the lepton momenta which we could actually measure at the LHC.

\section{Letting the particles decay}
\label{sec:decay}

In this section we consider what happens when the pair-produced particles decay: $Y\to l X$ and $\overline{Y}\to \overline{l}\,\overline{X}$.
Because $X$ and $\overline{X}$ are invisible to the detector, we are now in a position where we can no longer reconstruct the
final state: $\ct$, $x_1$, $x_2$ and therefore $\mathcal{A}_{ll}$ are no longer measurable. 

\begin{figure}[t]
\begin{center}
\includegraphics[width=7.5cm]{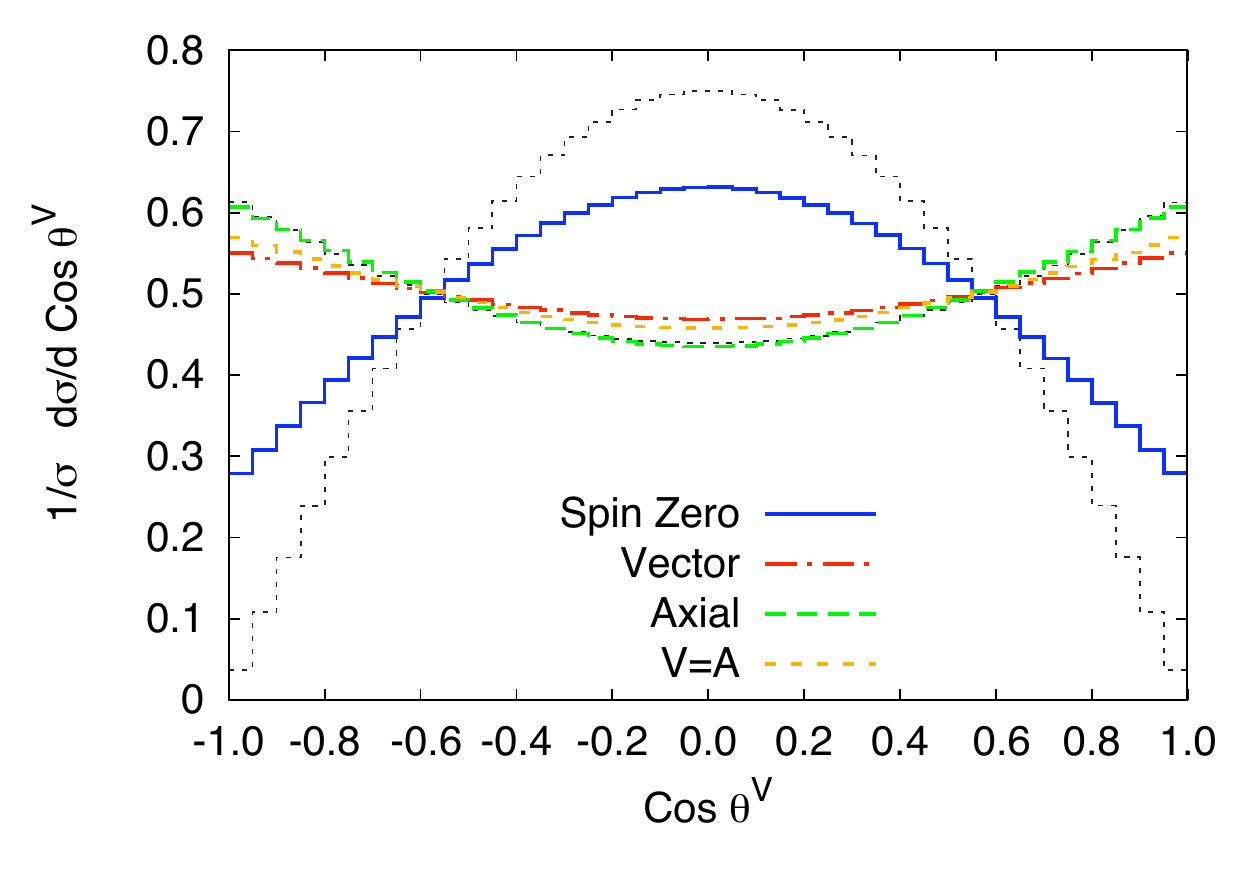}
\includegraphics[width=7.5cm]{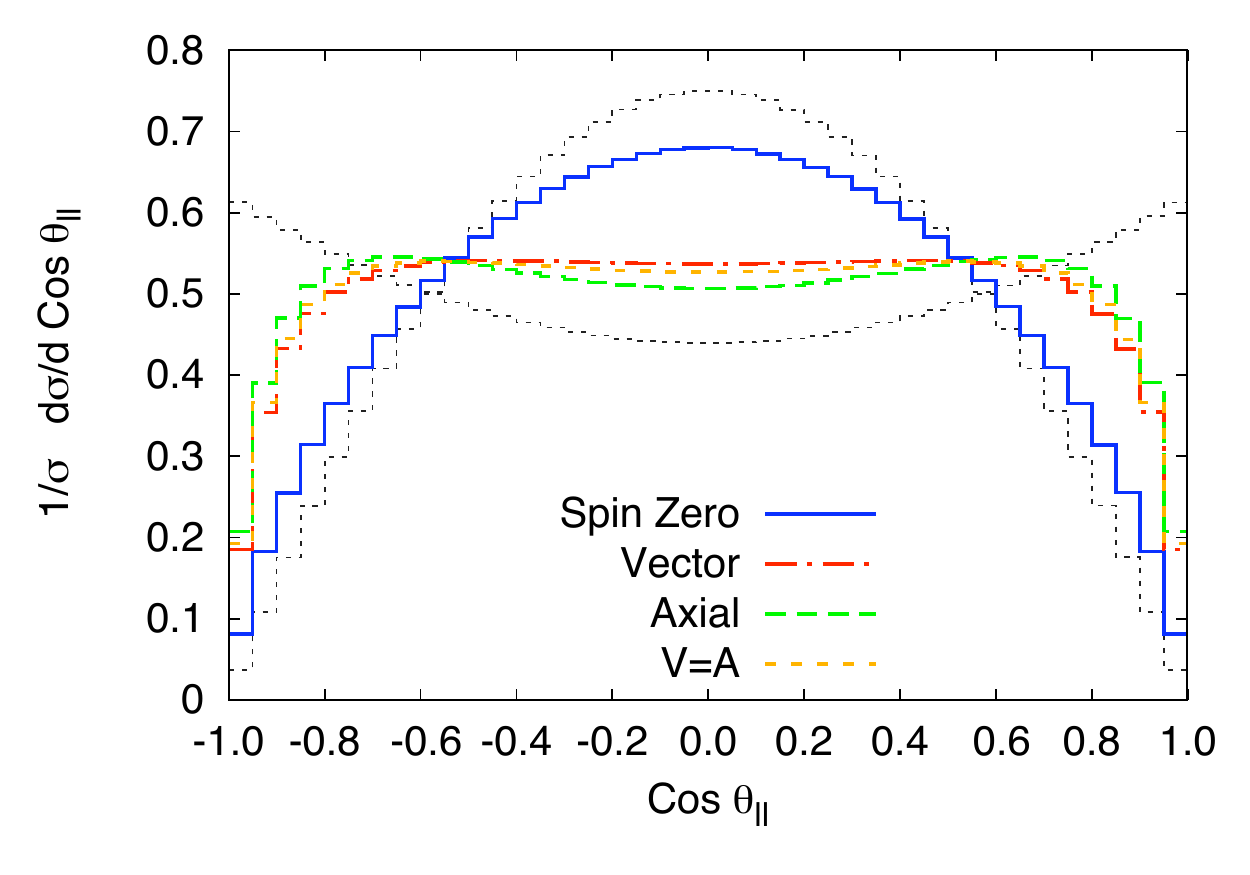}
\caption{Comparing the distribution of the variable introduced here, $\ctv$ (left), with that of $\ctll$ (right) for the different spin scenarios considered in the previous section. The faint, grey, dotted lines show the true $\ct$ curves -- those shown in figure~\protect\ref{cos} - for a spin zero $Y$, and for a spin half $Y$ with a purely vector coupling (in the left hand plot, this lies beneath the purely axial coupling (green) curve). We use the example masses for $Y$ and $Z$. The mass of $X$ has absolutely no impact on the shape of these distributions.}
\label{cosrec}
\end{center}
\end{figure}

We let the above decay happen isotropically in the rest
frame of $Y$, which means we take no spin correlations into account. To ignore spin correlations is to ignore an interesting way of
determining spin in these situations (if the correlations are there, $Y$ and $\overline{Y}$ are not spin zero), but here they only
have a small effect for two reasons. The first is that because all the final spin states are being summed over, the correlation is only
observable through an overall production polarisation of $Y$ and $\overline{Y}$, if one exists. Secondly, $Y$ and $\overline{Y}$ are produced with
a sizeable boost in the LAB frame -- this acts to wash out the effect of any lepton-lepton correlation.

\subsection{The visible $\ctv$ and the visible asymmetry $\Av$}
In reference \cite{Barr:2005dz}, Barr introduces a variable which approximates $\ct$ in the production and decay topology
described above. This variable $\ctll = \tanh(\Delta\eta_{ll}/2)$ is invariant under longitudinal boosts
as it is a function of $\Delta\eta_{ll}$, the pseudorapidity difference between the leptons. This is an attractive quality at the LHC where we do
not know the boost of the partonic centre of mass frame along the beam-axis.

In a similar fashion, we now introduce a new visible variable $\ctv$, which is defined as the cosine of the angle between $l$'s direction of motion and the beam-axis 
in the centre of mass frame of the two visible leptons. That is one takes the sum of lepton momenta $P = p_{l}+p_{\overline{l}}$, calculates
the invariant mass $m_{ll}=\sqrt{P^2}$ and boosts the lepton momenta into the frame in which $P=(m_{ll},0,0,0)$.
This is the reference frame in which the angle $\theta^V$ is defined. 
This new variable $\ctv$ coincides with $\ctll$ when the leptons have equal and opposite transverse
momenta. 

We plot the distribution of both of these visible variables in figure~\ref{cosrec} for $pp$ collisions at $14\,$TeV and using our example masses for $Y$
and $Z$. Now we need to select a mass for $X$ and we choose $M_X=500\,$GeV. The remarkable thing
is that we could choose any value of $M_X$ here ($M_X<M_Y$, of course) and the distributions shown in figure~\ref{cosrec} would stay exactly the same. 
We will discuss this rather interesting result below. In figure~\ref{cosrec} the true $\ct$ distributions are faintly shown and we can see that
$\ctv$ tracks these true curves better than $\ctll$, which suffers from a pseudorapidity `drag down' at large $|\ctll|$. However, the important  
point is how well the visible curves can be distinguished from one another. By eye one might expect $\ctv$ to perform better -- this is made into a 
statistical statement once detector effects have been taken into account in section~\ref{sec:coll}.

\begin{figure}[t]
\begin{center}
\includegraphics[width=6.3cm]{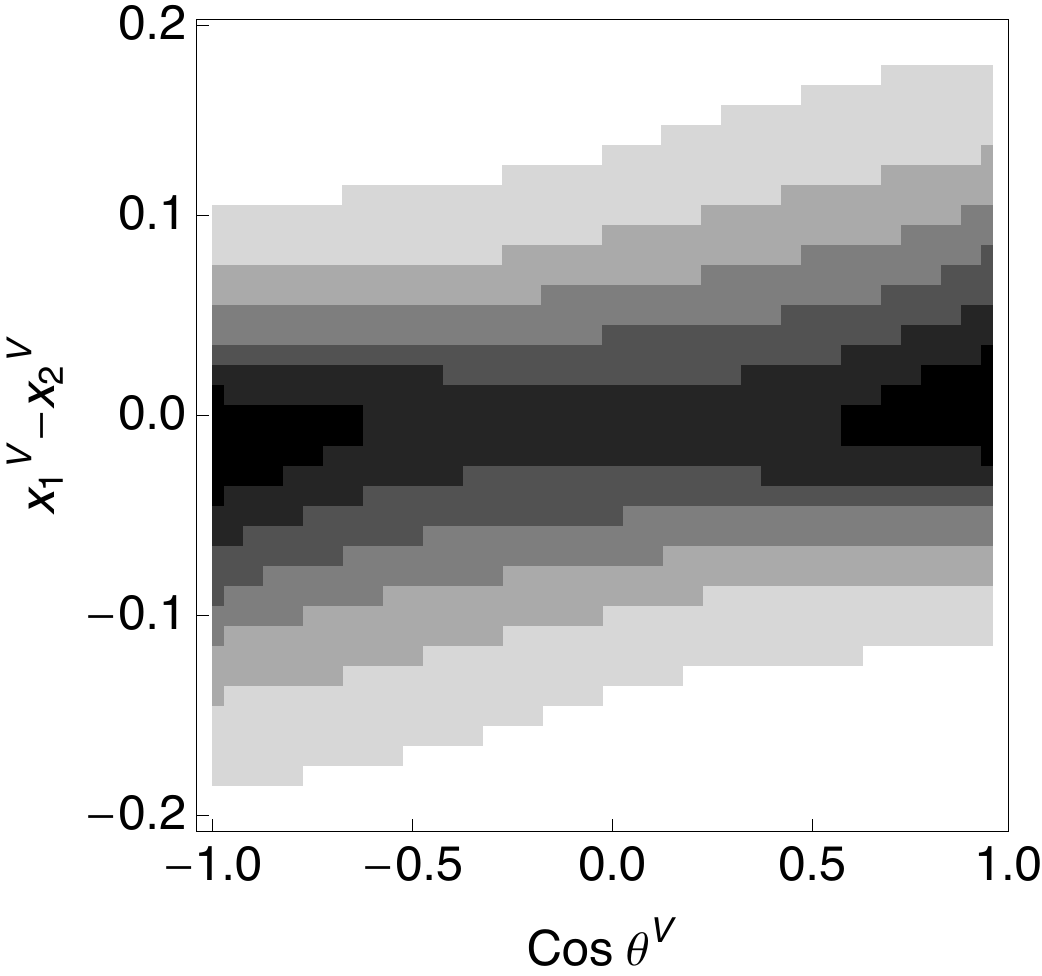}
\caption{An observable version of the asymmetry shown in figure~\protect\ref{cos} (spin half $Y$, $V=A=1$) is obtained by plotting events as a function of $\ctv$ and $x_1^V-x_2^V$.  Darker areas mean more events and the normalisation is arbitrary. We use the example masses for $Y$ and $Z$, and show the distribution for the choice $M_X=500\,$GeV. A larger value of $M_X$ does squash this distribution towards  $x_1^V-x_2^V=0$, but does not change the value of the asymmetry $\Av$.}
\label{All}
\end{center}
\end{figure}

To observe the asymmetry we introduce two other variables $x_1^V$ and $x_2^V$ 
\be
x_1^V = \frac{1}{\sqrt{S}}(E_{l}+E_{\overline{l}}+p^z_{l}+p^z_{\overline{l}}) \nonumber \\
x_2^V = \frac{1}{\sqrt{S}}(E_{l}+E_{\overline{l}}-p^z_{l}-p^z_{\overline{l}}) 
\label{xvis}
\ee
and define the asymmetry
\be
\Av = \frac{N^{x_1^V>x_2^V}_{\ctv>0}+N^{x_1^V<x_2^V}_{\ctv<0}-N^{x_1^V>x_2^V}_{\ctv<0}-N^{x_1^V<x_2^V}_{\ctv>0}}{N_{\text{Tot}}}\,.
\ee
We plot this visible asymmetry in the ($\ctv,x_1^V-x_2^V$) plane in figure~\ref{All}.

\subsection{Mass independence}

The variables we have just defined are exactly the same as the ones studied when we were pretending that we could measure $Y$ and
$\overline{Y}$'s momenta, except that the roles of $p_Y$ and $p_{\overline{Y}}$ have been taken up by $p_l$ and $p_{\overline{l}}$. This is no surprise -- it is
down to the fact that both $Y$ and $\overline{Y}$ have a relatively high boost in the LAB frame and so the leptons track their parents' direction well. 
We will begin by discussing this statement and will show that the angle between $l$ and $Y$'s direction of motion in any frame of reference
is independent of the mass of $X$. The way in which $\ctv$ and $\Av$ are independent of $M_X$, a perhaps slightly more surprising result, 
follows easily from this observation.

\begin{figure}[t]
\begin{center}
\includegraphics[width=7cm]{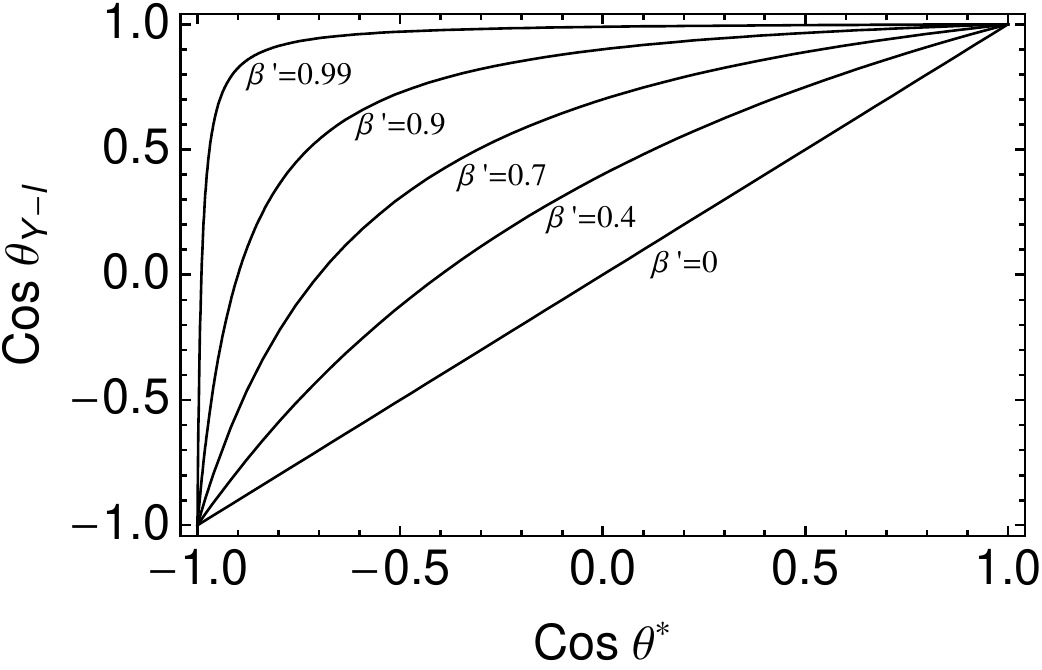}
\caption{ Plotting the relationship \protect \eqref{reln} between $\cos\theta_{Y-l}$ and $\cos\theta^*$ (both defined in the text) for various values of $\beta'$.
 A flat distribution in the centre of mass angle $\cos\theta^*$ leads to a distribution in $\cos\theta_{Y-l}$ which becomes
increasingly polarised towards $\cos\theta_{Y-l}=1$ as $Y$ becomes more boosted. }
\label{labvcm}
\end{center}
\end{figure}

Consider the decay of $Y$ in its own rest frame into $X$ and $l$. Here we write the momentum of $l$ as
\be
p^{\text{*}}_{l} = \left(\begin{array}{c}
E^*\\
0\\
E^*\sin\theta^*\\
E^*\ct^*
\end{array}\right)\,,
\ee
where $E^*=(M_Y^2-M_X^2)/2M_Y$. Now boost this to the frame in which $Y$ travels with momentum $p_z=M_Y\gamma'\beta'$ along the $z$ axis (the primes are
to distinguish these quantities from those used in the centre of mass frame of $Y$ and $\overline{Y}$) to obtain

\be
p_{l} = \left(\begin{array}{c}
\gamma'E^*(1+\beta' \ct^*)\\
0\\
E^*\sin\theta^*\\
\gamma'E^*(\beta'+\ct^*)
\end{array}\right)\,.
\ee
We can compute the cosine of the angle between $l$ and $Y$'s direction of motion
\be
\cos\theta_{Y-l} = \frac{\gamma'(\beta'+\ct^*)}{\big(\sin^2\theta^* + {\gamma'}^2(\beta'+\ct^*)^2\big)^{1/2}}\,.
\label{reln}
\ee
This result does not depend on our choice of the $z$-axis, and is true in a general frame of reference.
There are two things we wish to point out. Firstly, as $\beta'$ becomes large, the lepton really does start to track the direction
of its parent's motion very well, as illustrated in figure~\ref{labvcm}. This is how the variables introduced above
 can pick up on the information about the polar production angle of $Y$.
Secondly, $E^*$ has cancelled in this expression and so we see that this angle is independent of the mass
of $X$.

The way in which $E^*$ cancels here points us to the reason why $\ctv$ and $\Av$ are independent of $M_X$. The above result 
tells us that in any frame of reference, the momenta of $l$ and $\overline{l}$
take the form $p_l=E_l(1, \mathbf{\hat{n}})$ and $p_{\overline{l}}=E_{\overline{l}}(1, \mathbf{\hat{m}})$, where $\mathbf{\hat{n}}$
and $\mathbf{\hat{m}}$ are unit vectors which have a distribution unaffected by the value of $M_X$. All dependence on $M_X$ is contained within the overall factors
$E_l$ and $E_{\overline{l}}$ which are both proportional to $\Delta M^2 = (M_Y^2-M_X^2)$.
It follows then that \emph{any dimensionless observable in any reference frame must be independent of the mass of $X$ if the only dimensionfull inputs
are components of the leptons' four momenta}. The common factor of $\Delta M^2$ has to cancel. This completes the proof that $\ctv$ is independent of $M_X$. 
We also point out that $\ctll$, along with many other dimensionless variables which have been studied in the past such as
  $\Delta\phi_{ll}$, $\cos\theta_l-\cos\theta_{\overline{l}}$, etc. all enjoy this property too.

Turning to the asymmetry $\Av$, we now have a dimensionfull input which is not a lepton momentum component: $\sqrt{S}$. We can see from equation~\eqref{xvis}
 that $x_1^V-x_2^V$ is proportional to the mass difference $\Delta M^2$, so that the distribution shown in figure~\ref{All} gets squashed towards 
  $x_1^V-x_1^V=0$ for larger $M_X$. However, the value of $M_X$ has no effect on the sign of $x_1^V-x_2^V$, and so with the result for $\ctv$ we conclude
  that $\Av$ is also independent of $M_X$.

These are nice features for variables if they are to be used without any mass input. In the next section we will see how they can be used to determine
the spin of $Y$ once some idealised detector effects have been taken into account.

\section{Collider simulation and differently boosted $Y$s}
\label{sec:coll}

\begin{figure}[t]
\begin{center}
\includegraphics[width=7.5cm]{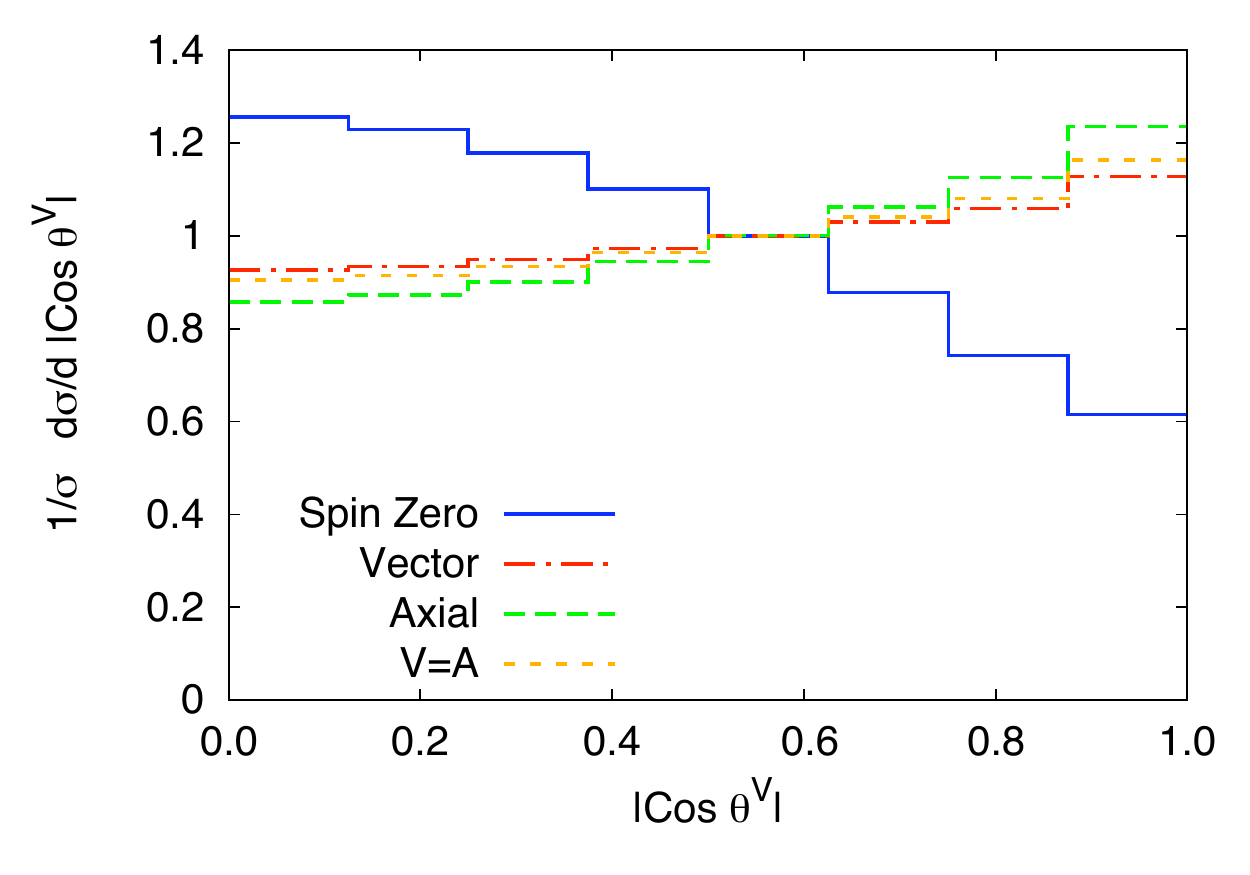}
\includegraphics[width=7.5cm]{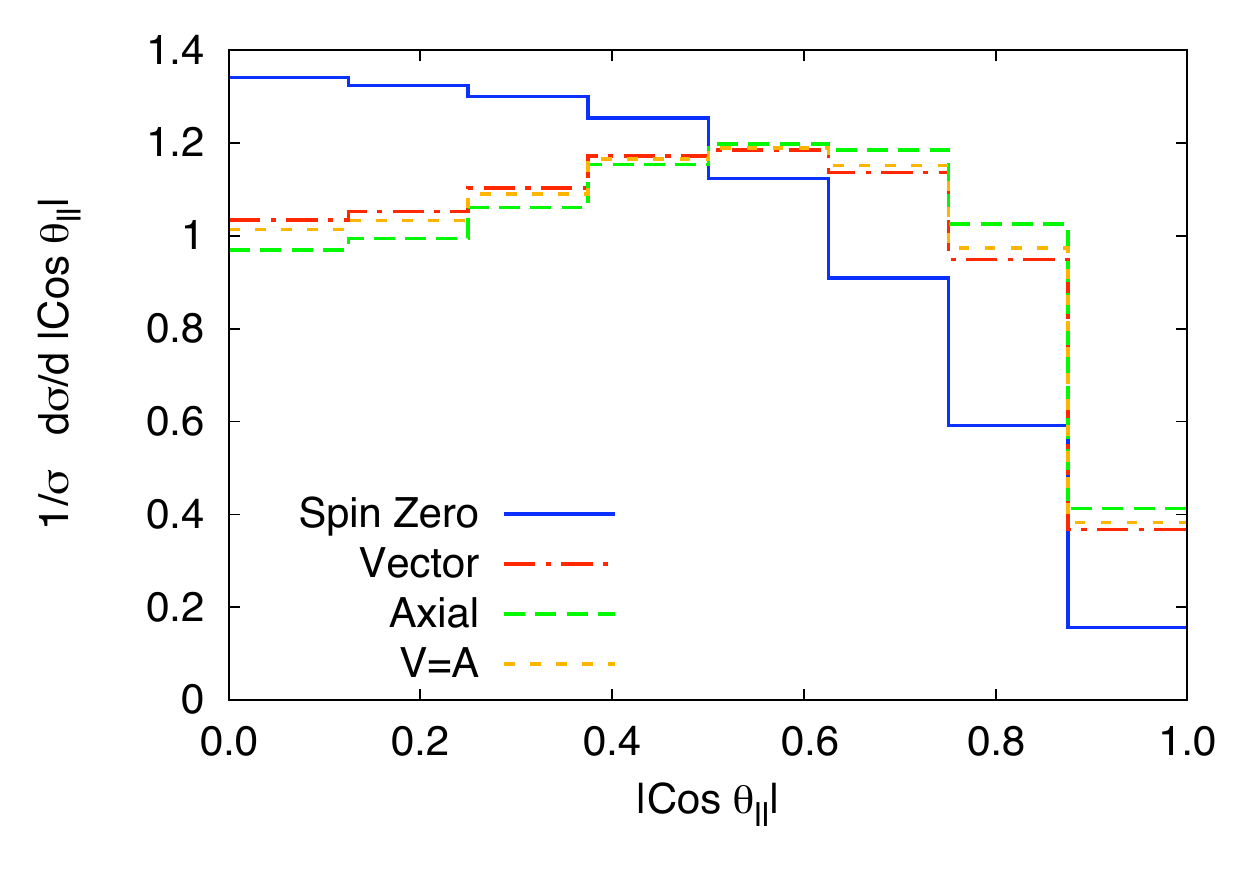}
\caption{ Comparing the normalised distributions of $\ctv$ (left) and $\ctll$ (right), after applying the kinematic cuts outlined in the text, for the four different spin scenarios considered throughout this paper. We use the example masses for $Y$ and $Z$, and here $X$ is not closely degenerate in mass with $Y$ -- so long as this is the case, these curves
are very insensitive to  $M_X$ (see text for details).   }
\label{coll}
\end{center}
\end{figure}

We begin by declaring that this collider simulation will be one skewed towards the theoretical end of the spectrum -- in particular we consider only partonic processes and
do not degrade the particles' momenta in any way. 
In reality many effects will lead to a smearing of the distributions -- for example, detector resolution and the issue of low $p_T$ initial state radiation\footnote{By this we mean immeasurable initial state radiation (ISR), in the sense that no hard jet is observed in the event.
It would be interesting to further study whether hard ISR in the event could be turned into something useful, as was done in \cite{Barr:2008ba}.} -- and we shall bear this in mind when making our conclusions. This approach is satisfactory because in \cite{Barr:2005dz} Barr carried out
 a realistic simulation far beyond the scope of this paper in the context of $\ctll$ and found that this variable could be used at the LHC for spin measurement 
 -- we can draw upon the similarities with this work below.
\begin{table}
\begin{center}
\begin{tabular}{|c|c|c|}
\hline
Process & $\sigma$ (fb) & $\Av$ \\
\hline
$WW$ & 0.98(1) & -0.044(1) \\
\hline
$ZZ$ & 0.46(1) & 0.22(1) \\
\hline
$WZ$ & 0.11(1) & 0.042(1) \\
\hline
\end{tabular}
\end{center}
\caption{Cross sections and values of $\Av$ for the main standard model background processes. Exactly two opposite sign, same flavour leptons ($e$ or $\mu$) satisfying
 the cuts described in the text were required, along with missing energy. These were produced at leading order using the POWHEG Box \protect \cite{Melia:2011tj} at the $14\,$TeV LHC. 
Renormalisation and factorisation scales were set to the sum of the two vector boson masses in the process. 
The error shown is statistical only. In the $ZZ$ case, only the two lepton, two neutrino final state was considered.}
\label{sm}
\end{table}
\begin{figure}[t]
\begin{center}
\includegraphics[width=7.5cm]{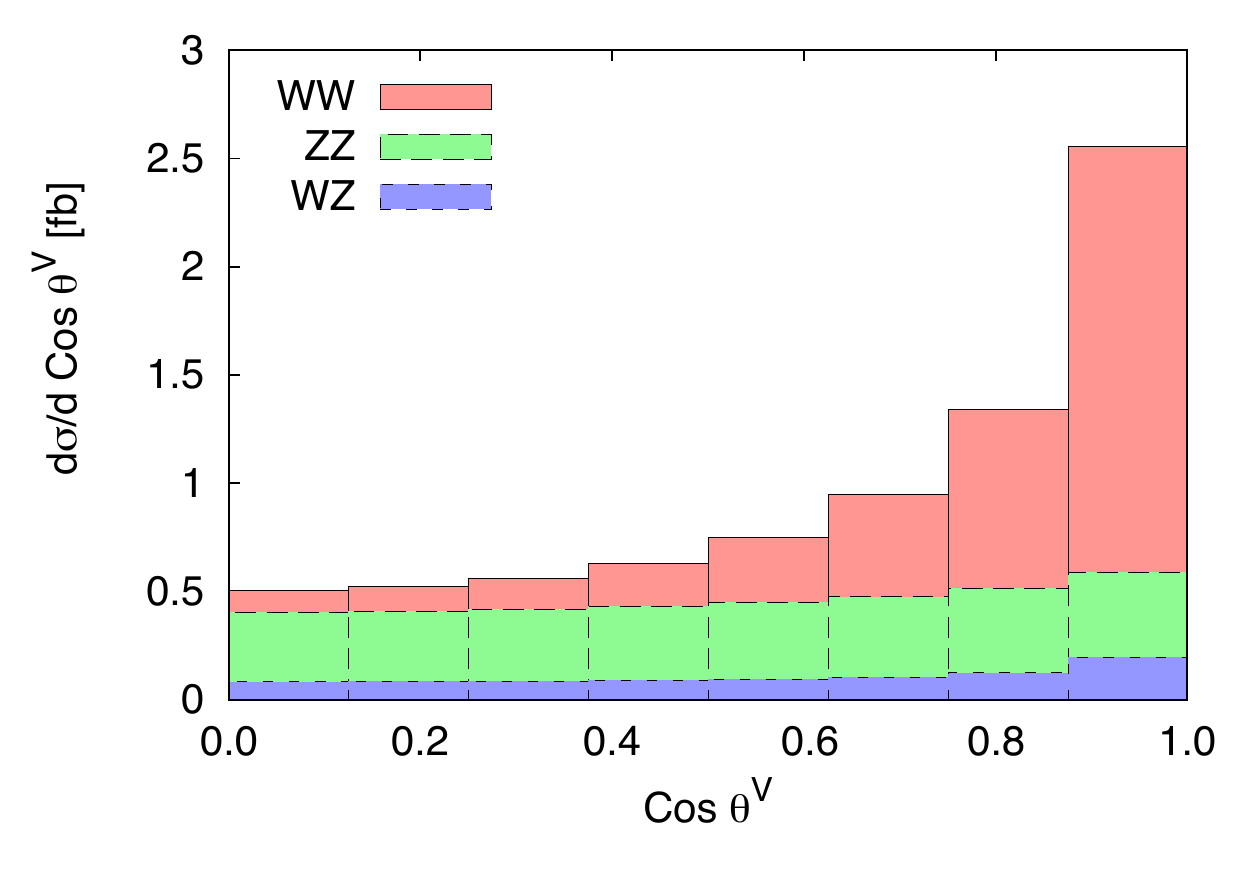}
\caption{The $\ctv$ distributions of the main standard model backgrounds in the two opposite sign, same flavour leptons plus missing energy channel, produced
in the POWHEG box for the $14\,$TeV LHC with the cuts described in the text.}
\label{back}
\end{center}
\end{figure}

We do however want to include detector effects which affect things 
at the parton level -- this involves making kinematic cuts and studying the impact these have on the signal and the standard model backgrounds to this process. The cuts
which we will employ are those of \cite{Barr:2005dz}: $p_{T,l_{\text{max}}}>40\,$GeV,  $p_{T,l_{\text{min}}}>30\,$GeV, where $l_{\text{max(min)}}$ is the lepton with the larger (smaller) transverse momentum;
 a cut on the rapidity of the leptons, $|\eta_l|<2.5$;  and a requirement for missing transverse momentum $p_{T,\text{miss}}>100\,$GeV. In addition,
 the following cuts are designed to remove the main standard model backgrounds: $M_{T2}>100\,$GeV and $M_{ll}>150\,$GeV\footnote{As the case we are considering
involves an on-shell $Z$, an even higher cut on $M_{T2}$ could in fact be used to reduce the $WW$ background further. }. The main backgrounds for this channel of 
opposite sign, same flavour leptons are $WW$ and $ZZ$ production (both decaying to $l^+l^-\nu\nu$), and we will also consider the $WZ$ production background.

In figure~\ref{coll} we plot normalised distributions for $\ctv$ and $\ctll$, after these cuts have been made, 
for a spin zero $Y$ and three types of spin half $Y$ (vector coupling, 
axial coupling, and for couplings $V=A=1$) at the $14\,$TeV LHC and with our example masses, $M_Y=800\,$GeV and $M_Z=2.5\,$TeV. Because we are now cutting on the lepton momenta, a slight dependence on the value of  the mass of $X$ is introduced and so we quote that the distributions shown are with $M_X=500\,$GeV. However, for
$M_X$ in the range $0-700\,$GeV these distributions only experience changes at the few percent level. For $M_X$ above
 this range, the signal cross section becomes very small as the cuts really start to have an effect;
in fact for $M_X=750\,$GeV the $M_{T2}$ cut entirely kills the signal\footnote{An animation which shows how
these distributions change as $M_X$ is varied from zero right up to the point at which the signal is killed by the cuts is available at \cite{myweb}.}. If the case where $X$ and $Y$ are closely degenerate in mass is realised at the LHC, 
it will be very difficult to see the signal process above the standard model background anyway. The value for the asymmetry $\Av$ for a spin half $Y$ with couplings
 $V=A=1$ is found to be $\Av=0.16(1)$, where the error
is purely statistical.

The main standard model background cross sections can be found in table~\ref{sm}, along with values for the asymmetry, $\Av$. Their distributions in $\ctv$ -- this time
\emph{not} normalised to unity -- are plotted in figure~\ref{back}. The importance of these backgrounds depends on the relative size of the beyond the 
standard model (BSM) process cross section. We include them to show their shape and magnitude. What was found in \cite{Barr:2005dz} was that they do not
pose a major problem for using $\ctll$ to determine spin for BSM cross sections of order a few tens of femtobarns (provided $X$ and $Y$ are not
closely degenerate in mass). We would expect this conclusion to hold true for $\ctv$ as well. Unfortunately, the standard model backgrounds, especially $ZZ$
production, yield rather large values of $\Av$. Since this has not been studied with realistic collider effects before, 
it is difficult to make a similar conclusion about the feasibility of observing
and using this asymmetry at the LHC for spin determination, especially without a BSM cross section in mind.

We now proceed to make a statistical statement about the power of $\ctv$ in determining the spin of $Y$, which comes down to being able to tell apart the different
distributions in figure~\ref{coll}. Following the method of reference \cite{Lester:2005je}, we estimate the number of events $N$ needed to disfavour one spin configuration
$S$ in favour of another, $T$, by a factor $R$ as
\be
N \sim \frac{\log R}{\text{KL}(T,S)}\,,
\label{kull}
\ee
where $\text{KL}(T,S)$ is the \textit{Kullback-Leibler distance} \cite{kldist} between the two curves. The results of this test are shown in table~\ref{nevts} for a value
of $R=1000$, where
the number of events needed for distinguishing a spin zero $Y$ from a spin half $Y$ using $\ctv$ are compared  with those needed using $\ctll$. We also show 
results for distinguishing an axially coupling spin half $Y$ from the other spin half cases (but not for telling apart a vector coupling and a mixed coupling $Y$ -- these
distributions lie almost on top of each other in figure~\ref{coll} and so will certainly be indistinguishable at the LHC).
 We shall comment on the number of events obtained in the context of spin before mass determination shortly, but for now we 
make two observations: firstly,
$\ctv$ performs around $20\%$ better than $\ctll$ in determining spin, and we would expect this to propagate through into a more realistic collider simulation; secondly,
using $\ctv$ to discriminate 
between the differently coupling spin half $Y$s requires enough events to render it an unrealistic 
approach at the LHC. 
Because of this, it would be very interesting if a more detailed and realistic detector 
study proved that $\Av$ could be used to give information about the couplings of a spin half $Y$.

\begin{table}
\begin{center}
\begin{tabular}{cc|c|c|c|}
\cline{3-5}
 & & \multicolumn{3}{|c|}{Variable Used} \\
\cline{1-5} 
 \multicolumn{1}{|c|}{T} & S & $\cos\theta$ & $\ctv$ & $\ctll$ \\
\hline 
 \multicolumn{1}{|c|}{Vector} & Spin Zero & 28 & 153 & 179 \\
 \multicolumn{1}{|c|}{Axial} & Spin Zero & 25 & 135 & 158 \\
 \multicolumn{1}{|c|}{$V=A$} & Spin Zero & 21 & 106 & 124 \\
 \hline
 \multicolumn{1}{|c|}{Vector} & Axial & 1126 & 4008 & 4870 \\
  \multicolumn{1}{|c|}{$V=A$} &Axial & 2487 & 8863 & 10794 \\
\hline
\end{tabular}
\end{center}
\caption{This table shows the number of events $N$ needed to discredit the spin configuration $S$ in favour of spin $T$ 
using the distributions of the variables $\ctv$ and $\ctll$ in figure~\protect\ref{coll}. For reference, the number of events needed using
(the immeasurable) $\ct$ are included. The top half of the table shows comparisons between spin zero and spin half cases, and
the bottom half shows some comparisons between different coupling spin half cases.}
\label{nevts}
\end{table}

\begin{figure}[t]
\begin{center}
\includegraphics[width=10cm]{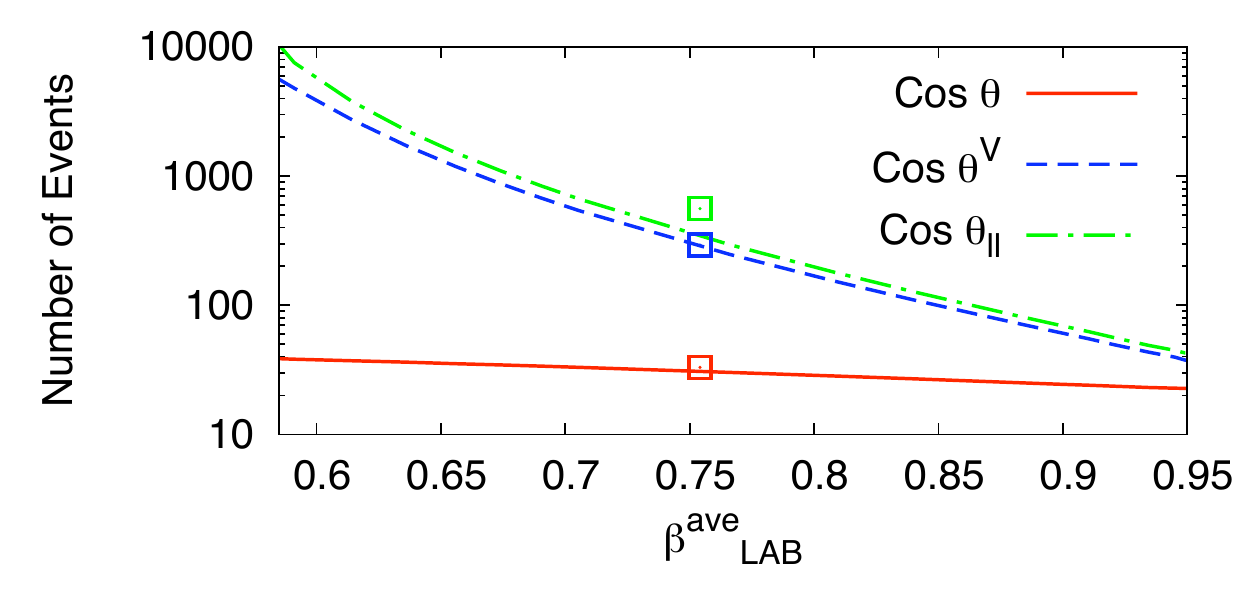}
\caption{ This plot shows the number of events needed to discredit the spin zero $Y$ hypotheses in favour of a spin half (purely vector coupling) $Y$ as a function of $\beta_{~~\text{\tiny LAB}}^{\text{ave}}$ -- the average boost of $Y$ in the LAB frame. This parameter is affected by the relative values of $M_Z$ and $M_Y$, and is what is relevant for this method of spin determination (see text for details of how $\beta_{~~\text{\tiny LAB}}^{\text{ave}}$ is varied here). Curves for $\ctv$, $\ctll$ and (the immeasurable) $\ct$ are compared. The results for the case when $Z=Z^0$, the standard model weak boson, are plotted as open squares. }
\label{beta}
\end{center}
\end{figure}

We move on to address the question of how things change when we pick different values for the mass of $Y$ and $Z$. The relevant parameter 
which is affected by these masses is the boost of $Y$ in the LAB frame, and in figure~\ref{beta} we show how the number of 
events needed to disfavour a spin zero $Y$ hypothesis in favour of a vector coupling spin half $Y$ changes as the average value of this parameter, 
$\beta^{\text{ave}}_{~\text{\tiny LAB}}$, varies. We choose to do this by keeping the mass of $Y$ fixed at $800\,$GeV
 and varying the mass of $Z$. The
lowest value of $\beta^{\text{ave}}_{~\text{\tiny LAB}}$ is where we approach the limit of threshold production of $Y\overline{Y}$ in the $Z$ rest frame. Shown also are results
for the case when $Z$ is the standard model $Z^0$ -- these points lie off the other curves because in this case the shape of the $\beta$ spectrum is different, 
owing to the fact that we are no longer considering an on-shell resonance decaying to $Y\overline{Y}$. As  $\beta^{\text{ave}}_{~\text{\tiny LAB}}$
tends towards one, we see the behaviour we would expect from figure~\ref{labvcm} -- the number of events using the three different variables all tend towards the same value. This is because
both $\ctv$ and $\ctll$ approximate ever more closely the true $\ct$ distribution, owing to the fact that the decay products from an increasingly relativistic $Y$ will follow ever more
closely the direction in which $Y$ travels. On the other hand, as $\beta^{\text{ave}}_{~\text{\tiny LAB}}$ decreases, the number of events needed to determine $Y$'s spin
using $\ctv$ and $\ctll$ rises exponentially, but we note that at the same time, $\ctv$ becomes increasingly better at determining spin than $\ctll$. In the case where $Z=Z^0$, there is
a factor of two gain in statistics in using $\ctv$. The number of events needed diverge at threshold production due to the fact we considered an isotropic decay of $Y$.
An animation which shows the changes taking place in the $\ctv$ and $\ctll$ distributions as $\beta^{\text{ave}}_{~\text{\tiny LAB}}$ 
is varied is also available at the website \cite{myweb}.

Is it possible to determine spin before mass here? Given what was said in the introduction about the measurement of mass, this question really reduces
to whether it would be possible to determine spin in this topology during the course of an LHC run. 
Realistic collider effects will certainly increase the number of events 
indicated in table~\ref{nevts} and figure~\ref{beta}, 
but so long as $Y$ and $\overline{Y}$ are produced with a boost greater than $\sim0.75$ then one might
expect to be able to make a spin measurement (spin zero or spin half) with a few hundred events. 
This would require the BSM cross section to be at least around the $10\,$fb level. 
Furthermore,
$X$ and $Y$ can't be too closely degenerate in mass if the signal is not to be lost in the standard model background.
These conclusions are backed up by what was found in the particular cases studied in \cite{Barr:2005dz}. Finally, with reference to early spin measurement, 
we point out that the value of $R=1000$ chosen in equation~\eqref{kull} could be relaxed somewhat which would give a (less certain) indication of the spin of $Y$
with fewer events.
 
\section{Discussion and outlook}
\label{sec:conc}

In this paper we have introduced two new variables which can be used for spin determination at the LHC: $\ctv$ and $\Av$. We considered the $s$-channel  pair-production of particles $Y$ and
$\overline{Y}$ by a spin one particle $Z$, and looked at the topology where $Y\overline{Y}\to l X\overline{l}\,\overline{X}$. We showed how $\ctv$ can give information about
 the distribution of the polar production angle of $Y$, illustrating its use in
 discriminating between the case of a spin zero and a spin half $Y$. We found that in this theoretical study $\ctv$ possesses more 
statistical power than the related variable $\ctll$, and would expect this to propagate through when more realistic collider effects are taken into account, as they 
were in the original study of $\ctll$ \cite{Barr:2005dz}. 
The asymmetry
$\Av$ which we introduced has the potential to identify a spin half $Y$ which has both vector and axial couplings to the $Z$ production particle, something which cannot be
done using $\ctv$ alone. 
Until further study, we reserve judgement on the usefulness of this variable for spin determination at the LHC --
without specifying
a particular value for the cross section of the new physics it is unclear how much of a role the standard model background would play in obscuring such a measurement,
and a simulation of realistic collider effects is required here as well. 
It would be very interesting to see if a measurement of this asymmetry were possible at the LHC outside of the context
of spin determination, especially
as the standard model predicts a non-zero value for it in the two leptons plus missing energy channel.

We proved that neither of these variables depend on the mass of the invisible decay product $X$, and pointed out that a number of other variables also
enjoy this property -- we are not aware that this was known before. 
The event topology is one in which mass determination is particularly difficult and so we considered the question of what could be said about spin without
the knowledge of any new particle masses. 
Because of their independence of the mass of $X$, the question of whether these variables can be used to measure spin could be assessed on the basis of one parameter -- the
boost at which $Y$ and $\overline{Y}$ are produced in the LAB frame.
We found that so long as they are reasonably well boosted, discrimination 
between the two spin assignments in this set-up using $\ctv$ becomes possible after a few hundred events, before any of the particle masses are known.

\section*{Acknowledgements}

I would like to thank Alan Barr, Chris Lester, John March-Russell, Kirill Melnikov, Graham Ross, Markus Schulze and Giulia Zanderighi for their helpful 
discussions and explanations, 
comments on the manuscript, and enthusiasm during the course of this work.
This research was supported by a UK Science and Technology Facilities 
Council (STFC) studentship.

\bibliography{sb4m}
\bibliographystyle{JHEP}

\end{document}